\documentstyle[preprint,aps,psfig]{revtex}

\newcommand{\nea}{\nearrow}
\newcommand{\sea}{\searrow}
\newcommand{\nwa}{\nwarrow}
\newcommand{\swa}{\swarrow}
\newcommand{\J}{{\cal J}}

\begin{document}
\bibliographystyle{prsty}
%\twocolumn[\hsize\textwidth\columnwidth\hsize\csname@twocolumnfalse\endcsname
%%%%%%%%%%%%%%%%%%%%%%%%%%%%%%%%%%%%%%%%%%%%%%%%%%%%%%%%%%%%%%%%%%%%%%%%%%%%

\title{Microscopic Theory of Magnetic Phase Transitions in HoNi$_2$B$_2$C}
\author{A. Amici$^{1}$ and P. Thalmeier$^{2}$}
\address{
$^1$Max-Planck-Institut f\"{u}r Physik komplexer Systeme, 01187 Dresden
Germany,\\
$^2$Max-Planck-Institut f\"{u}r chemiche Physik fester Stoffe,
01187 Dresden, Germany
}
\date{\today}
\maketitle

\begin{abstract}

We present a microscopic theory for the low temperature metamagnetic 
phase diagram of HoNi$_2$B$_2$C that agrees well with 
experiments.
For the same model we determined the
zero field ground state as a function of temperature and find the $c$-axis
commensurate to incommensurate transition in the expected
temperature range.
The complex behaviour of the system
originates from the competition between the crystalline electric field and the
Ruderman-Kittel-Kasuya-Yosida interaction, whose effective form is obtained.
No essential influence of superconductivity has to be invoked to understand 
the magnetic phase diagram of this material.

\end{abstract}

\vspace{.5cm}
\newpage
\vfill
%] %matches %\twocolumn

%%%%%%%%%%%%%%%%%%%%%%%%%%%%%%%%%%%%%%%%%%%%%%%%%%%%%%%%%%%%%%%%%%%%%%%%%%%%
%{\small PACS: 61.43.-j; 61.43.Hv; 02.50.+s.}

The recent interest in HoNi$_2$B$_2$C and similar borocarbide compounds 
is motivated by the possibility of a detailed study of the 
mutual interaction between superconductivity (SC) \cite{nat94}
and magnetic order (MO) \cite{canf94,gri94} coexisting 
in a few of these materials as bulk properties.
The superconducting critical temperature
for HoNi$_2$B$_2$C is T$_c$ = 8 K \cite{nat94} and the upper 
critical field H$_{c2}^{sc}$ is about 2 $\sim$ 3.5 kG \cite{lin95}.
Temperature dependent measurements
show a pronounced anomaly of H$_{c2}^{sc}$(T)
around 5 K which nearly leads the material to reentrance into
the normal state \cite{canf94,eis94}. In the same temperature range
several magnetically ordered structures are observed
in the Ho f-electrons sub-lattice:
a commensurate (C) antiferromagnetic phase below 5 K \cite{canf94,gri94},
an incommensurate (IC) $c$-axis complex spiral state (5 K $<$ T $<$ 6 K) 
\cite{gri94,gold94,hill96} and
an $a$-axis IC modulation in a narrow range of temperature
around 5.5 K \cite{gold94,gg95}.
Although no satisfactory theory is available presently,
many experiments point to a correlation between
this complex magnetic phase diagram and the SC anomalies \cite{rat96,sch97}.
Much insight on the magnetism of this
material can be gained from a number of experiments
reporting metamagnetic transitions at low temperature
and fields higher than H$_{c2}^{sc}$ \cite{canf94,cho96}.
In particular the detailed anisotropic metamagnetic 
phase diagram at T = 2 K presented in Ref. \cite{canf97},
can be used to extract many features of the magnetic interaction.
In this work we propose a realistic microscopic model for
the Ho $4f$-electrons sub-system of HoNi$_2$B$_2$C which reproduces
the main features of the low temperature metamagnetic phase diagram
as well as the zero field sequence of phases as function of
temperature.
It is important to note that many physical elements
need to be included in the present theoretical description
to have reasonable agreement with experiments
and likely a model of a comparable complexity
is needed to treat the mutual interaction between SC and MO.

The coexistence and weak coupling of
the two phenomena of SC and MO is due to
the different degrees of localisation of electrons
in the borocarbides. LDA
calculations show that the conduction
band is composed mainly of Ni $3d$-electrons \cite{matt94}, which
undergo the superconducting transition. On the other hand magnetic properties
are related to the well localised electrons in the incomplete
$4f$-shell of Ho. The exchange interaction
between these two electron systems is mediated by
the small fraction of Ho $6s$ and $5d$ character in the conduction band.
As far as magnetic properties are concerned the conduction electrons 
can be eliminated in a standard way leading to an effective RKKY
exchange interaction among the stable Ho $4f$ moments.
The appropriate Hamiltonian is then given by \cite{jen91}:
\begin{equation}
{\cal H} = \sum_i [{\cal H}_{cf}({\bf J}_i) -
	\mbox{\boldmath $\mu$}_i \cdot {\bf B} ] -
	\frac{1}{2} \sum_{ij} \J(i,j)\, {\bf J}_i \cdot {\bf J}_j
\label{ham3d}
\end{equation}
This Hamiltonian include the crystal electric field (CEF) single
ion part ${\cal H}_{cf}({\bf J}_i)$ expressed in terms of the 
total angular momentum {\bf J}$_i$, the Zeeman interaction between
the local magnetic induction {\bf B} and the
magnetic moment {\boldmath $\mu$}$_i$ = $\mu_B$g{\bf J}$_i$
($\mu_B$ is the Bohr magneton and $g = \frac{5}{4}$ 
the gyromagnetic ratio for Ho),
and the effective RKKY exchange interaction.
The direct dipole-dipole interaction is not relevant in borocarbides
\cite{kul97} and this allows us to use a magnetic induction
field {\bf B} independent of the position.
The CEF single ion Hamiltonian we use is the one extracted from 
neutron diffraction experiments in
Ref. \cite{gas96} and contains no adjustable parameter.
Its ground state is a $\Gamma_4$ singlet and the first exited states are
a $\Gamma_5^*$ doublet at 0.15 meV from the ground state
and a $\Gamma_1$ singlet at 0.32 meV.
The other 13 CEF states have much higher excitation energies ($>$ 10 meV) and
their matrix elements with the low energy quartet is very small,
therefore they may be neglected in the whole range of temperature and fields
which we explored.
Regarding the magnetic interaction no previous knowledge
is available about the RKKY function $\J(i,j)$ and an important aim of this work
is to obtain a realistic model for it.

Since we are dealing with a three dimensional system of
large angular momenta (J$_{Ho}$ = 8) it is possible to treat its Hamiltonian
at a mean field (MF) level. 
Introducing the mean thermal value $\left<{\bf J}_i\right>$ and neglecting
the terms containing two sites fluctuations it is possible to decouple
the dynamics of the different sites. The 
single ion MF Hamiltonian is then given by \cite{jen91}:
\begin{equation}
{\cal H}_{mf}(i) = {\cal H}_{cf}({\bf J}_i) -
	{\bf J}_i \cdot {\bf B}_i^{e}
	+ {\cal E}
%	+ \frac{1}{2} \left<{\bf J}_i\right> \sum_j \J(i,j)\left<{\bf J}_j\right>
\label{mf3d}
\end{equation}
in which {\bf B}$_i^{e}$ = $(\mu_Bg{\bf B} + \sum_{j} \J(i,j)\, 
\left<{\bf J}_j\right>)$ is the effective molecular field and
${\cal E} = \frac{1}{2} \left<{\bf J}_i\right> 
\sum_j \J(i,j)\left<{\bf J}_j\right>$.
The diagonalisation of
this single site Hamiltonian can be easily achieved
numerically. The single ion Gibbs free energy density F$^M_i$
and the corresponding average angular momentum 
$\left<{\bf J}_i\right>$
can be computed as functions of {\bf B}$_i^e$ and T. This leads to the
self consistent MF equations for the $\left<{\bf J}_i\right>$
which can be solved iteratively. 
Two experimental evidences can be used to 
reduce the number of independent sites
whose MF magnetisation should be computed.
First of all, almost all the observed
structures in this compound share the property of
ferromagnetic alignment in the $ab$ plane.
The only exception is the $a$-axis modulation whose structure
is not yet clear. However it appears
not to coexist microscopically with the $c$-axis ones \cite{hill96}
and we will neglect it.
Therefore we impose ferromagnetic alignment in the plane from the beginning
thus reducing the calculation to one dimension. Then the RKKY interaction
have only $c$-dependence and, taking the magnetic 
moment in the site 0 as the reference,
it may be parametrised with the help of
%\begin{equation}
${\cal J}_{i} = \sum_{\{j_i\}} {\cal J}(0,j_i)$,
%\end{equation}
with j$_i$ running on all the sites of the i$^{th}$ plane.
The $\J_i$ with $i \ge 1$ are the interaction of the reference moment with
those in the neighbouring layers and $\J_0$ is the interaction with the
other moments in the same plane.
The second simplification is related to the CEF
structure, in fact the $c$-axis is a very hard
direction in the whole range of temperature we are interested in,
therefore we force the moments to lie in
the $ab$ plane neglecting their small out-of-plane component.

In order to establish the actual MF ground configuration for the 
$\left<{\bf J}_i\right>$ out of the possible stable ones 
we introduce the Helmholtz free energy 
(HFE) density given by:
\begin{equation}
  F({\bf r}) = F^M({\bf B}) + \frac{B^2}{8\pi} - 
\frac{{\bf H} \cdot {\bf B}}
{4\pi}
= F^M({\bf B}) - 2\pi M^2 - \frac{H^2}{8\pi} 
\end{equation}
with {\bf B}({\bf r}) = 
{\bf H}({\bf r}) + 4$\pi${\bf M}({\bf r}) \cite{lan84}.
The integral of F({\bf r}) over all space
is the proper thermodynamic function to be minimised when
external fields are kept constant \cite{lan84}.
The magnetisation contribution ($\sim$ 10 kG) is not negligible with respect
to the typical external field (4-25 kG). However the full problem
of solving for {\bf M}({\bf r}) in a finite sample with a given geometry
is beyond our purpose. This is indeed a typical problem
in thermodynamics of magnetic materials \cite{lan84} 
whose most practical solution is
to assume cylindrical symmetry around the external field {\bf H} in order
to eliminate the spatial dependence of both {\bf B} and {\bf M}.
The homogeneous magnetisation of the sample is calculated as
{\bf M} = $\frac{g\mu_B}{V_c}\overline{\left<{\bf J}_i\right>}$ where 
V$_c$ = 65 \AA$^3$
is the volume of the unit cell and the bar indicates the average
on all ions. Similarly 
the contribution of the magnetisation to the HFE per unit cell
can be written as $2\pi M^2V$ = 
$\frac{2\pi (g\mu_B)^2}{V_c}\overline{\left<{\bf J}_i\right>}^2$.

Until now the actual RKKY interaction among magnetic 
moments remained unspecified. In order to achieve 
a convenient parametrisation for it we make an extensive use of
the T = 2K magnetisation data in Ref. \cite{canf97}.
From the clear presence of flat magnetisation plateaux
and from the value of the magnetisation
as a function of the angle
we argue
that the magnetic moments of the ions are almost
at the saturation value J$_s$ = J$_{Ho}$ = 8 and they are locked in one of the
four equivalent $\left<110\right>$ in-plane easy directions \cite{canf97}.
The sharp metamagnetic transitions are then due to
first order transitions among different arrangement of the 
moments along the $c$-axis. We assume that the relevant phases in the
field-angle phase diagram are the easiest
commensurate structures with the observed magnetisation
(they are listed in the upper part of table \ref{tab1}).
We notice that all the phases in table \ref{tab1} may be represented
in a chain of six unit cells with periodic boundary
conditions. This is important because the Fourier
transform of the interaction for a system with six sites have only four
free parameter, namely $\J_i$ with i = 0, 1, 2, 3. Any further
Fourier component may be removed with a renormalisation of these
parameters (i.e. the $\J_6$ is equivalent to $\J_0$).
Neglecting the CEF energy and entropy it is possible
to calculate analytically the HFE for each configuration.
Some of these energies
are listed in table \ref{tab1}.
From the two $\theta$ = 0$^\circ$ metamagnetic transitions it
is possible to establish two conditions for the parametrisation of the
RKKY interaction. Following the convention of Canfield 
{\em et. al.} \cite{canf97}
we call the metamagnetic transition fields H$_{c1}$ = 4.1 kG and H$_{c2}$
= 11.1 kG (this value is somewhat higher than the 10.6 kG given in Ref. 
\cite{canf97}), they should
not be confused with the superconducting critical fields.
Imposing the energy of the AF2 and AF3 phases to
be equal at H$_{c1}$ and the ones of AF3 and P at H$_{c2}$
we obtain the following relations:
\begin{eqnarray}
{\cal J}_{1}&=& - {\cal J}_2 - \frac{g \mu_B}{2 J_s}H_{c2} 
- \frac{2}{3} E_M
\\
%{\cal J}_{2}&=& {\cal J^*} \\
{\cal J}_{3}&=& {\cal J}_2 + \frac{g \mu_B}{6 J_s}(H_{c2}
- H_{c1}) 
+ \frac{1}{6} E_M
\nonumber
\end{eqnarray}
where $E_M = \frac{2\pi (g\mu_B)^2}{V_c}$ = 8.1 10$^{-3}$ meV
comes from the contribution of the magnetisation to the HFE.
They assure that the relative stability of the three phases
for $\theta$ = 0$^\circ$ is the observed one.
In order to use the easiest possible model we set $\J_3$ = 0,
which implies $\J_1$ = $-$8.0 10$^{-3}$ meV and $\J_2$ = $-$2.4 10$^{-3}$ meV.
The final freedom in the model is the parameter $\J_0$ which
works basically as a self interaction and cannot be extracted
starting from the energy differences among magnetic
structures.
We use $\J_0$ = 4.8 10$^{-3}$ meV in order to have the
transition between the paramagnetic and the incommensurate state 
at the experimental value T = 6 K \cite{lin95}.
Moreover this choice make the internal molecular
field large enough ($\sim$ 15 kG) to maintain the moments close to the
saturation regime, as required.

\begin{figure}
%\vspace{7.0cm}
\centerline{\psfig{file=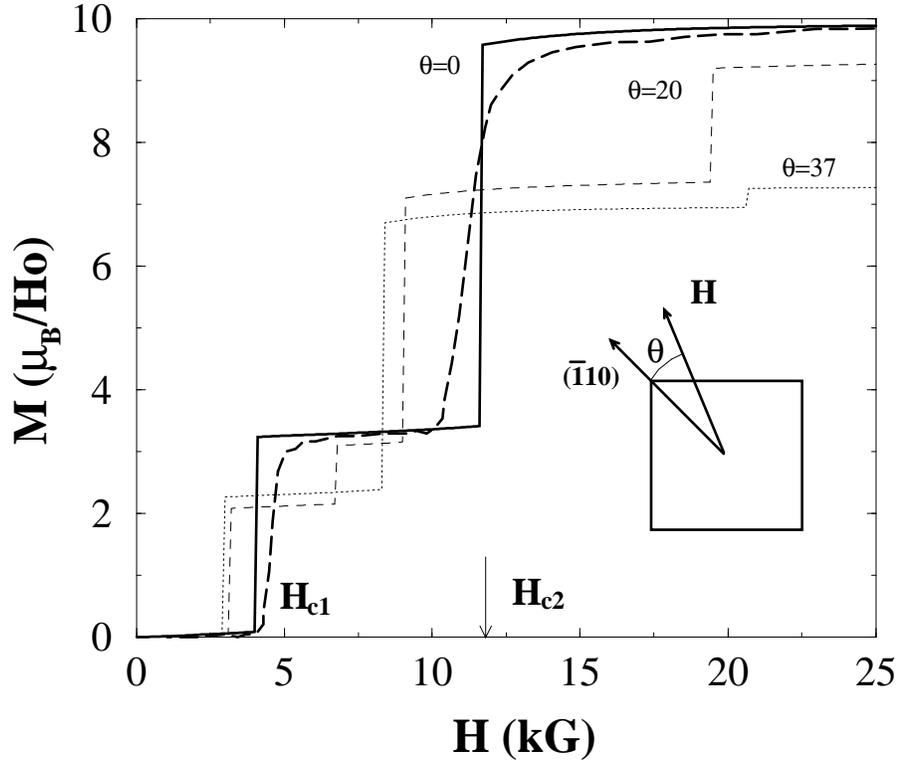,height=11.0cm}}
\protect\caption{Magnetisation vs. magnetic field for
some representative angles $\theta$ between the field and the closest 
$\left<110\right>$
direction. For $\theta$=0$^\circ$ the thick dotted line represent experimental
data taken from Canfield {\em et al} [12]. and the thick continuous line give
results of our calculation. The two experimental parameters entering our
model are the transition fields of the two metamagnetic transition
H$_{c1}$ = 4.1 kG and H$_{c2}$ = 11.1 kG. 
This figure should be compared with fig. 1 (a) of [12]}
\label{fig1}
\end{figure}

\begin{figure}
%\vspace{7.0cm}
\centerline{\psfig{file=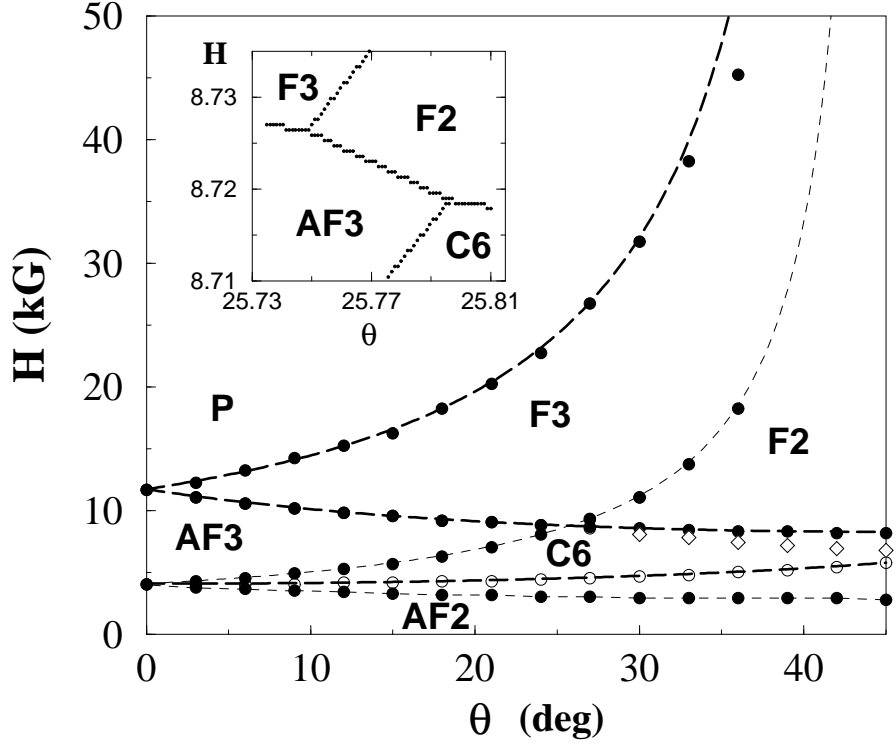,height=11.0cm}}
\protect\caption{Field vs. angle metamagnetic phase diagram for T = 2 K.
The filled dots are calculated within our model and
indicate the phase transitions where a sensible jump 
in the magnetisation is seen
($\Delta$M $>$ .2 $\mu_B$/Ho). The thick dashed lines are
given in [12] as the best fit of the experimental data.
Their functional form is H$_{c3}^\circ$/sin(45$^\circ - \theta$),
H$_{c2}^\circ$/cos(45$^\circ - \theta$)
and H$_{c1}^\circ$/cos$\theta$. The values we used for the proportionality
constants are
H$_{c3}^\circ$ = 
H$_{c2}^\circ$ = 8.4 kG and H$_{c1}^\circ$ = 4.1 kG. The two additional
thin curves refer to the phase boundaries not yet observed in the experiments.
%Both of them are expected to have hysteresis since the symmetry of the
%phases are different. Hysteresis is not expected to be large
%for other transitions.
Empty symbols refer to the stability of the AF3 phase with respect to AF2
(circles) and F2 (diamonds).
The apparent tetracritical point in the phase diagram
is composed of two very close usual tricritical points (inset).
}
\label{phases}
\end{figure}

After the model has been defined and all its parameters fixed we
present now the numerical results of the complete self consistent MF
calculation for the field-angle phase diagram at T = 2 K.
The starting values for $\left<{\bf J}_i\right>$ in the iteration
algorithm are a set of random numbers
and the possible periods allowed are in the range
from one to nineteen planes.
Typical magnetisation curves are evaluated and they 
are shown in fig. \ref{fig1},
they refer to experimental geometries with different angles $\theta$ 
between the external field {\bf H} and the closest magnetic easy axis 
$\left<110\right>$.
The resulting metamagnetic phase diagram is presented
in fig. \ref{phases}.
It contains all the phases in the table \ref{tab1} 
and it agrees remarkably well with the experimental one.
In particular all the transition lines show simple trigonometrical
dependence as a function of 
the angle $\theta$ as well as the corresponding magnetisation values
at the plateaux.
In all the regions where the stable phases are AF2, AF3, F3 and P
we observe remarkable quantitative agreement with the experimental data
in Ref. \cite{canf97}.
The main qualitative differences between
our model and experiments is the presence of two additional phases,
the F2 and the C6, whose magnetisation is different from the reported ones.
However the strongest disagreement is in the region of
relatively low field and high angles where experimentally the strongest
hysteresis is found. 
Another minor point is the absence of direct
AF3-P transition for $\theta \ne$ 0$^\circ$. Experimentally the direct
transition is observed for small angles
up to $\theta$ = 6$^\circ$ in the Ho compounds,
but seems to have a much wider range in the similar phase diagram of 
DyNi$_2$B$_2$C \cite{canf98}.
We would like to stress however that it is not possible to improve
this phase diagram simply refining the parameters for the RKKY
interaction. For example allowing for $\J_3 \ne$ 0 it is possible
to stabilise at low field an additional phase,
which is a distorted helix structure with wave length five,
but no change appears in the phase transitions among the other
phases.
In addition, as explained before, the effect of
further effective interactions with 
neighbouring planes beyond the third one may be eliminated by proper
renormalisation of the $\J_i$ with i $<$ 3.
Therefore if the two phases F2 and C6 are not observed in
the experiments we have to conclude that additional interactions
(i.e. magnetoelastic couplings) have to play
a role in the stability of magnetic phases in
HoNi$_2$B$_2$C. 

Starting from the same model it is possible to
analyse the zero field behaviour as a function of
temperature. In principle this requires some attention since the magnetic
system is now embedded in a superconducting material.
This implies important changes in the {\bf q} $\sim$ 0
region of the Fourier transform of the RKKY function \cite{andboh}, but leaves
the relevant {\bf q} $\sim \pi$ region almost unchanged.
Relying on this fact and on experiments on doped non-superconducting materials
such as HoNi$_{2-x}$Co$_x$B$_2$C which show a magnetic behaviour very
similar to the undoped superconducting one \cite{lynn96,all97}, 
we will use our purely magnetic
model for the description of the zero field phase diagram.
At MF level the second order phase transition
between the paramagnetic state and an ordered structure is expected
to occur a the {\bf Q} vector for which the $\J$({\bf q})
has its maximum. In our model this correspond to
Q$_c$ = 0.78 $\pi$ in the $\left<001\right>$ direction, not
far from the experimental value Q$_{exp}$ = 0.91$\pi$ \cite{hill96}.
The helical state is
preferred with respect to the longitudinal modulated structure due to the
$ab$ easy plane for the moments. This truly incommensurate
structure can be the ground state of the system only as long
as the average moment per ion is small enough, i.e. 
close to the transition temperature.
Lowering the temperature the ordered state develops and
the CEF part of the HFE, proportional to fourth
and the sixth powers
in $\left<{\bf J}_i\right>$, force the structure to find
a commensurate compromise.
Because in the self consistent MF treatment it is not
possible to treat at the same time truly incommensurate structures
and the CEF, we cannot observe the actual C-IC transition. However
at a temperature of T = 5.5 K is observed a first order transition from
AF2 to a helical state of
wave length 17, where the moments point no longer 
only along the easy directions.
This is the clear indication that the
RKKY energy starts to become dominant with respect to the CEF potential and
drives the system into a state whose wave number {\bf Q} is closer to the
maximum of the RKKY function.
To obtain better quantitative agreement for
the ordering wave number {\em and} for the temperature interval in
which the incommensurate state is stable, the function $\J$({\bf q})
has to be refined in the Q$_c$ region by including
further parameters.

In conclusion, we presented a microscopic model for the 
the rare earth borocarbide system HoNi$_2$B$_2$C which explains
the main reported features of the anisotropic magnetic
and temperature phase diagrams.
The minimal model to achieve a semiquantitative description of the complex
magnetic behaviour of the system needs to include realistic
CEF and effective RKKY interaction among the planes.
On the other hand no influence of SC needs to be included in the
determination of the magnetic structures observed.

\vspace{.5cm}
We would like to thank W. Henggeler {\em et al.} for the data on the CEF
states. A.A. would also like to thank M. Laad, P. De Los Rios and B. Canals.
This work was performed under DFG Sonderforschungsbereich 463.

\begin{table}
%\begin{centering}
\begin{tabular}{clc}
%\hline
Phase & Structure       & HFE per site \\
\hline
 P         & $\nwa$                         &
$- J_s^2(\J_1 + \J_2 + \J_3 + E_M) -J_s H_\parallel$ \\
 AF2       & $\nwa\sea$                     & 
$J_s^2(\J_1 - \J_2 + \J_3)$       \\
 AF2$'$       & $\nea\swa$                     & 
$J_s^2(\J_1 - \J_2 + \J_3)$       \\
 AF3       & $\nwa\sea\nwa$                 &
$[J_s^2(\J_1 + \J_2 - 3\J_3 - E_M/3) - J_sH_\parallel]/3
%- J_s^2 E_M/9
$  \\
 AF3$'$       & $\nwa\nea\swa$                 &
$[J_s^2(\J_1 + \J_2 - 3\J_3 - E_M/3) - J_sH_\parallel]/3
%- J_s^2 E_M/9
$  \\
 F3        &  $\nwa\nea\nwa$                &
%$[-J_s^2(\J_1 + \J_2 + 3\J_3) 
%- 2J_sH_\parallel-H_\perp]/3 - 5/9 J_s^2 E_M$ 
\\
\hline
 F2        &  $\nwa\nea$                    & 
%$- J_s^2\J_2 - J_s(H_\parallel+H_\perp)/2 - 1/18 J_s^2 E_M$  
\\
 C6        &  $\nwa\sea\nwa \nea\swa\nea$   & 
%$[J_s^2(4\J_1 - 2\J_2) -J_s(H_\parallel- H_\perp)]/6 
%- 1/18 J_s^2 E_M $
\\
\end{tabular}
\protect\caption{Stacking sequence of ferromagnetically ordered
$ab$-planes along the $c$-axis for the phases found in the T = 2 K 
magnetic anisotropic phase diagram in fig. 2. 
At low temperature the CEF forces the moments
to lie in one of the four easy direction $\left<110\right>$ 
indicated by arrows.
The external field forms an angle $\theta$ with the (-1,1,0) 
($\nwa$) direction.
The third column is the relevant part of the HFE per site, the
term $-J_{s}^2\frac{\J_0}{2} - \frac{H^2}{8\pi}$ may be added to have the total
HFE.
H$_\parallel$ = g$\mu_B$H cos($\theta$) is the projection of the
field along the easy axis and
% H$_\perp$ = g$\mu_B$H sin($\theta$),
all other symbols are explained in the text.
We only give the HFE for the phases needed to compute the relations
(4).}
\label{tab1}
%\end{centering}
\end{table}

\end{document}